\documentclass[prb,aps,floats,twocolumn,showpacs]{revtex4}
\usepackage{amsmath}
\usepackage{graphicx}
\usepackage{epsfig}
\usepackage{float}

\newcommand{\bea}{\begin{eqnarray}}
\newcommand{\eea}{\end{eqnarray}}
\newcommand{\ket}{\rangle}
\newcommand{\bra}{\langle}

\newcommand{\be}{\begin{equation}}
\newcommand{\ee}{\end{equation}}

\begin{document}
 \title{ Adiabatic quantum pumping of a desired ratio of spin current to charge current}
 \author{Sungjun Kim, Kunal K. Das \footnote{Present Address: Department of Physics, Fordham University, Bronx,
New York 10458}, and Ari Mizel}
 \affiliation{  Department of Physics, The Pennsylvania State University, University Park, Pennsylvania 16802, USA }

\begin{abstract}
We present a prescription for generating pure spin current or spin
selective current, based on adiabatic quantum pumping in a
tight-binding model of a one dimensional conductor.  A formula for
the instantaneous pumped current is derived  without introducing
the scattering matrix. Our calculations indicate that some pumping
cycles produce the maximum value 2 of pumped spin while
others reverse the direction of current as a result of small
alterations of the pumping cycle.  We find pumping cycles which
produce essentially any ratio of spin current to charge current.
\end{abstract}


\maketitle

\section{ Introduction }
The field of spintronics offers a vision of electronics that utilizes
carrier spin in addition to carrier charge \cite{Samarth}.  The rich
potential of carrier spin for applications ranges from non-volatile
devices to quantum computation.  In order to realize this potential,
however, it is essential to develop effective tools for the
manipulation and transport of spin.  Adiabatic quantum pumping is a
mechanism that can transport charge at zero
bias\cite{Brouwer,Avron,Entin,Buttiker}. As a result of cycling two or more physical parameters that characterize a one-dimensional conductor, charge carriers get ``swallowed'' down the conductor
like food down a throat, comprising a direct current.
This method can deliver precise currents and requires no voltage bias.
Recently, it has been shown 
that adiabatic quantum pumping in the presence of a magnetic field can
also generate a spin current\cite{Sharma,Mucciolo,Watson,Aono,Wei}.
For some fortuitous choices of experimental parameters, it has even
been possible to generate a spin current without any charge current,
which is termed a ``pure'' spin current \cite{Mucciolo,Watson,Wei}.
This is a promising development, and one wonders if it is possible to
establish complete control over both the amount of spin and the amount
of charge pumped per cycle.

A device for generating a pure spin current with improved control
appears in Ref. \onlinecite{Aono}, in which Zeeman energy is chosen as
one of the adiabatic pumping parameters.  There is no need to make
a fortuitous choice of parameters in this device --
when the minimum and maximum Zeeman energies involved in the pumping
cycle are equal in magnitude but opposite in sign, a pure spin
current arises.  However, if the maximum value of Zeeman energy is
not equal and opposite the minimum value, various combinations of spin
current and charge current arise in a way that is not easily controlled.

In this paper, we introduce a flexible approach to adiabatic
pumping in which essentially any composition of spin/charge
current can be chosen as desired.  The scheme utilizes
``generalized'' pumping parameters each of which depends on more
than one physical parameter.  With generalized pumping parameters,
many different physical processes map to the same path in pumping
parameter space. The result is greatly improved control over the
pumping products.

For instance, a pure spin current can be generated in the
following way.  In adiabatic quantum pumping, carriers are transported
with each cycling of the pumping parameters.  By reversing the
direction of the pumping cycle trajectory, one reverses the direction
of the current flow.  With generalized parameters, it is possible to
make the spin-up pumping parameters traverse the exact same trajectory as
the spin-down pumping parameters, but in the opposite direction.  The
result is that spin-up carriers get pumped in one direction and
spin-down carriers in the other direction, leading to zero net
transport of charge but non-zero spin current.  This technique can be
applied to any desired cycle in parameter space, so that the amount
of spin pumped per cycle can be set by judicious choice of the
trajectory in parameter space.

Generalized pumping parameters also enable selective spin
pumping wherein the current consists only of spin-up carriers or only
of spin-down carriers.  One spin's parameters traverse a degenerate
cycle that pumps no charge, while the other spin's parameters traverse
a productive cycle.  This selective spin pumping is a valuable tool;
by combining and repeating spin selective pumping, it is possible to
generate any rational proportion of spin current to charge
current.

Finally, in addition to pure spin pumping and spin selective
pumping, we consider a family of cycles which produce arbitrary
ratios of spin current to charge current after exactly one cycle.
Unlike the schemes mentioned above, the correct cycle in this case
cannot be fixed in a deterministic way; trial and error is necessary. 
However, we present an argument that an appropriate
cycle generally exists. Moreover given a cycle that produces a
given ratio of spin current to charge current, we also show how to
traverse a cycle that produces the inverse ratio.

The rest of the paper includes a presentation of our model
Hamiltonian, in sec. II, and a derivation of the pumped current.
Section III. discusses the generation of a pure spin current. In
Sec. IV, we present a method for selective spin pumping of just
one spin orientation. A means of generating arbitrary ratios of
spin current and charge current is proposed in Sec. V, and we
conclude with Sec. VI.


\begin{figure}[b]
\epsfig{figure=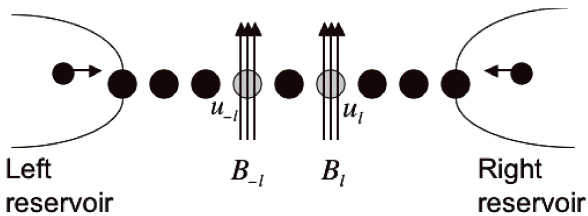,width=3.2 in}
\caption{ Gray dots located at $-l$ and $l$ sites represent impurities
on a wire; the impurities are characterized by potential barriers
$u_{-l}$,$u_{l}$. Localized magnetic fields $B_{-l}$,$B_{l}$ are
applied on those impurities.
\label{Fig:model}}
\end{figure}


\section{ Model Hamiltonian and Pumped Current }

We consider transport through a one-dimensional channel of sites,
schematically depicted in Fig. \ref{Fig:model}.  The following is our
model Hamiltonian
\bea
H &=& H_{0} + V_{1} + V_{2} \notag \\ H_{0} &=&
-J \sum_{n , \sigma } a_{n+1 \sigma}^{\dagger} a_{n \sigma} + a_{n
\sigma}^{\dagger} a_{n+1 \sigma} \notag \\ V_{1} &=& \sum_{ \sigma }
u_{-l} \: n_{-l \sigma} + u_{l} \: n_{ l \sigma } \notag \\
V_{2} &=& \sum_{ \sigma } - \sigma E^{Z}_{-l} \: n_{-l \sigma} - \sigma E^{Z}_{l} \: n_{ l \sigma }. \label{eq: H}
\eea

In the Hamiltonian,$-J$ is the nearest neighbor hopping amplitude,
$a_{n\sigma}^{\dagger}$ is the electron creation operator at site $n$
for spin $\sigma$ and $n_{l \sigma} = a_{l\sigma}^{\dagger} a_{
l\sigma}$ is the number operator at site $l$ for spin $\sigma$.  The
first term $H_{0}$ is the Hamiltonian of carriers in a homogeneous
channel and $V_{1}$ is the impurity potential.  The impurities are
characterized by on-site energies $ \{ u_{-l},u_{l} \} $, which
simulate potential barriers at sites $-l$ and $l$.  The $V_{2}$ term
contains the Zeeman spin energies $E^{Z}_{-l} = g \mu_{B} B_{-l}$,
$E^{Z}_{l} = g \mu_{B} B_{l}$ for carriers in the localized external
magnetic fields at sites $-l$ and $l$.  We assume that all four
experimental parameters in Fig. \ref{Fig:model} can be tuned
precisely; naturally, this would be challenging to realize in the
laboratory.  We set spin $|\sigma |$ equal to 1 instead of $1/2$.  We
assume that the four quantities $ \{ u_{-l}, u_{l}, E^{Z}_{-l},
E^{Z}_{l} \}$ are slowly varying time-dependent parameters. We define
generalized spin-dependent pumping parameters $ \{ X_{-l \sigma },
X_{l \sigma } \}$.  \bea X_{-l \sigma} &=& ( u_{-l} - \sigma
E^{Z}_{-l} ) / J \notag \\ X_{ l \sigma} &=& ( u_{ l} - \sigma E^{Z}_{
l} ) / J \label{eq: X} \eea in terms of which the Hamiltonian becomes
\bea H &=& H_{0} + V \notag \\ H_{0} &=& - J \sum_{n , \sigma } a_{n+1
\sigma}^{\dagger} a_{n \sigma} + a_{n \sigma}^{\dagger} a_{n+1 \sigma}
\notag \\ V &=& J \sum_{ \sigma } X_{-l \sigma } \: n_{-l \sigma} +
X_{l \sigma} \: n_{ l \sigma }. \label{eq: finalH} \eea
For mathematical convenience the following are set to unity $\hbar
= e = \: lattice \: spacing \: = 1$. By defining the
spin-dependent parameters  $ \{ X_{-l \sigma }, X_{l \sigma } \}$,
we make explicit the fact that spin-up carriers and spin-down
carriers are separately controllable.

Instantaneous scattering states $| \chi_{p \sigma} \ket$ of the
Hamiltonian are obtained by ignoring the time-dependence of the
pumping parameters and using the Lippmann-Schwinger
equation\cite{Economou} 
\be | \chi_{p \sigma} \ket = [ 1 + G
(E_{p})  V ] \: c^{\dag}_{p \sigma} | 0 \ket \ee
where $c^{\dag}_{p \sigma}$ is the carrier creation operator for
energy $E_{p} = -2 J \cos p$ and spin $\sigma$, and $G(E_{p})
= 1 /( E_{p} - H + i \eta )$ is the retarded full Green's function, $\eta = 0^{+}$. 
The time-variation of the potential is
adiabatic if it is slow compared to the dwell time of a carrier in
the scattering region \cite{Buttiker-Landauer}. 
The time-dependence is then restored to first order by adiabatic corrections \cite{Entin}

\be
| \phi_{p \sigma } \ket = | \chi_{p \sigma} \ket - i \: G (E_{p}) \: | \dot{\chi }_{p \sigma } \ket \label{eq:firstorder}
\ee  

In terms of these first order scattering states
(\ref{eq:firstorder}), the instantaneous pumped current associated
with spin $\sigma$ is
 \bea j_{\sigma} (n) &=& \int_{ - \infty}^{\infty} dE \:
F(E) \biggl[ -2 J \: \Im \int_{- \pi}^{\pi} \frac{ dp }{ 2 \pi }
\: \delta (
E - E_{p} ) \times \notag \\
 & & \: \bra n \sigma | \phi_{p \sigma } \ket \bra n+1 \sigma | \phi_{p \sigma } \ket^{\ast} \biggr] \label{eq: current}
\eea where $\Im$ indicates imaginary part, $F(E)$ is the Fermi distribution function, and $ | n
\sigma \ket = a^{\dag}_{n \sigma }  | 0 \ket $. 

First the integral over $p$ is evaluated

{\small
\bea
 & & -2 J \: \Im \int_{- \pi}^{\pi} \frac{ dp }{ 2 \pi } \: \delta ( E - E_{p} ) 
             \: \bra n \sigma | \phi_{p \sigma} \ket \bra n+1 \sigma | \phi_{p \sigma} \ket^{\ast}     \\   
 &\approx& - 2 J^{2} \: \Im \sum_{m = \pm  l } \:  \dot{ X }_{m \sigma}  \int_{-\pi}^{\pi} \frac{ dp }{ 2 \pi }  
      \: \delta ( E - E_{p} ) \times   \notag  \\  
 & &       \: \biggl[
      i  \bra n +1 \sigma | \:G^2(E_{p})  \: | m \sigma \ket^{\ast}  \bra n  \sigma |\chi_{p \sigma} \ket 
         \bra m \sigma | \chi_{p \sigma} \ket^{\ast}    \notag    \\ 
 & &     - i  \bra n \sigma | \:G^2(E_{p}) \: | m \sigma \ket 
     \: \bra m \sigma  | \chi_{p \sigma} \ket  \bra n + 1 \sigma  |\chi_{p \sigma} \ket^{\ast} \biggr]  \label{eq: firstordercalculation }   
 \eea} 
where the identity $ | \dot{ \chi }_{ p \sigma}  \ket = G(E_{p}) \dot{V} | \chi_{p \sigma} \ket $ has been used\cite{Entin}. 
Eq.\ (\ref{eq: firstordercalculation }) is evaluated using the identity
{\small
\bea
 & & \int_{-\pi}^{\pi} \frac{ dp }{ 2 \pi }  
     \delta ( E - E_{p} ) \: \bra n \sigma |\chi_{p \sigma } \ket \bra m \sigma  | \chi_{p \sigma } \ket^{\ast}  \notag  \\  
 &=&  - \frac{1}{ \pi } \frac{1}{2i}  
     \biggl[  G ( n \sigma  ,m \sigma ; E ) - G^{\ast} ( n \sigma ,m \sigma ;E ) \biggr]   \label{eq: crucialidentity }   
\eea}  
where $ G ( n \sigma  ,m \sigma ; E ) = \bra n \sigma |  G(E)  | m \sigma \ket$. 
The result is 
{\small
\bea
 & &  \frac{J^{2}}{\pi} \: \Im \sum_{m = \pm  l} \:  \dot{ X }_{m \sigma } \:  
      \biggl[
        \bra n +1 \sigma  |  G^{2}(E)  | m \sigma  \ket^{\ast}  \times  \notag   \\ 
 & &       \bigl[  G ( n \sigma  ,m \sigma ;E ) - G^{\ast} ( n \sigma  ,m \sigma ;E ) \bigr] \notag  \\  
 & &     - \bra n \sigma  |  G^{2}(E)   | m \sigma  \ket  \times  \notag   \\ 
 & &   \bigl[  G ( m \sigma ,n+1 \sigma ;E ) - G^{\ast} ( m \sigma ,n+1 \sigma  ;E)\bigr] 
       \biggr].    \label{eq: closeform }      
\eea}  

Eq.\ (\ref{eq: closeform }) can be simplified 
if we use the fact that the 1-D Green's function takes a plane-wave form at large distances\cite{Entin}
{\small
\bea
G (n+1 \sigma ,m \sigma ; E_{k}) &=& e^{ik} G (n \sigma ,m \sigma ;E_{k} ) \quad for \: \: n \rightarrow  \infty. \notag  \\
& &  \label{eq: asymptoticidentity }
\eea}
The asymptotic condition  $ n \rightarrow   \infty $ means the observation point is far away from the scattering center 
and is located on the right side. 
Inserting this into Eq.\ (\ref{eq: closeform }) produces
{\small
\bea
& &  - \frac{J^{2}}{\pi} \: \Im \sum_{m = \pm  l } \:  \dot{ X }_{m \sigma}    
          \: \partial_{E} \bigl[ G ( n\sigma ,m \sigma ;E ) G^{\ast} ( n +1 \sigma , m \sigma ;E ) \bigr].  \notag \\
& &  \label{eq: derivativeform }
\eea}  
At the last step of the calculation, we use the identity 
$ \bra n \sigma | G^2(E) | m  \sigma \ket = - \partial_{E} G (n  \sigma ,m  \sigma ;E ) $. 
This identity is derived by expanding $G(E)$ in a basis of energy eigenstates of $H$,
$G(E) = \sum_{\mu} \frac{ |E_{\mu} \ket \bra E_{\mu} | }{ E - E_{\mu} + i \eta}$.    
 The integration over energies in Eq.\ (\ref{eq: current}) can be performed with the result (\ref{eq: derivativeform }) 
at zero temperature where the Fermi distribution function is a step function
{\small
\bea
  & &  j_{ \sigma } ( n )  =    
        - \frac{J^{2} }{\pi} \sum_{m = \pm  l}  \dot{ X }_{m \sigma }  \times  \notag  \\
  & &	  \Im \bigl[ G ( n \sigma  ,m \sigma ;E ) G^{\ast} ( n +1 \sigma  , m \sigma ;E ) \bigr] |_{E=E_{F}}   
           \: for \: n  \rightarrow \infty.   \notag    \\                 
  & &    \label{eq: finalclosedform } 
 \eea}
This closed form for the instantaneous pumped current in
terms of retarded full Green's functions is one of the main results of this paper.  
In contrast to most treatments of quantum pumping, our derivation does not introduce the scattering matrix.
(However, see e.g. Ref. \onlinecite{Liliana}.) The identity (\ref{eq: crucialidentity }) is crucial in our derivation.

To evaluate the pumped current associated with the model Hamiltonian
(\ref{eq: H}), we need to compute the full Green's function.  The full
Green's function can be expressed in terms of the free Green's
function by the algebraic identity\cite{Economou} $G(E) =
G_{0}(E) + G(E) V G_{0}(E)$, where $ G_{0}(E) = 1 / (E
- H_{0} + i \eta )$. Eq.\ (\ref{eq: finalclosedform }) is evaluated to be

{\small
\bea
 \lefteqn{j_{\sigma } (n)  
	=   - \frac{J}{ \pi} \sum_{m = \pm  l} \:  \dot{ X }_{m \sigma} \times }  \notag  \\ 
	& &     \biggl[    
		\frac{1}{ 2} \bigl[ f (-m \sigma ) f^{\ast} (-m \sigma )  + h (-m \sigma ) h^{\ast} (-m \sigma ) \bigr] 
		     \Im [ G_{0} ( 0 \sigma ,0 \sigma ;E ) ]  \notag   \\ 
        & &       +  \Re [ f (-m \sigma ) h^{\ast} (-m \sigma ) ]  \Im [ G_{0} ( m \sigma ,-m \sigma ;E ) ]  \notag  \\ 
  	& &       -  sign (m) \Im [ f (-m \sigma ) h^{\ast} (-m \sigma ) ] \Re [ G_{0} ( m \sigma ,-m \sigma ;E ) ]
     \biggr] \bigg{|}_{E=E_{F}}  	\label{eq: evaluableform } 	 
		 \eea}	
where

{\small
  \bea
  f (m \sigma ) &=& \frac{ 1 - J X_{m \sigma } G_{0} ( 0 \sigma  ,0 \sigma ;E  ) }{ Z_{m \sigma } } \notag \\  
  h (m \sigma ) &=& \frac{ J X_{m \sigma } G_{0} ( m \sigma ,-m \sigma ; E ) }{ Z_{m \sigma } } \notag \\ 
  Z_{m \sigma } &=& 1 - J [ X_{-m \sigma } + X_{m \sigma } ] G_{0} ( 0 \sigma ,0\sigma ;E )  \notag \\ 
              & & + J^{2} X_{-m \sigma } X_{m \sigma } G_{0}^{2} ( 0 \sigma ,0 \sigma ; E )   \notag  \\
	      & & - J^{2} X_{-m \sigma } X_{m \sigma } G_{0}^{2} ( m \sigma  ,-m \sigma ; E ).  \notag 
  \eea}

Eq.\ (\ref{eq: evaluableform }) can be evaluated since the explicit expression for the free Green's function \cite{Economou}
is just $ G_{0} (n \sigma ,m \sigma ; E_{k} ) =  e^{ ik | n-m| } / 2 i J  \sin k $. 
The pumped charge associated with spin $\sigma$ after one pumping cycle 
is obtained by integrating Eq.\ (\ref{eq: evaluableform }) with respect to time 
\bea
 q_{ \sigma }   &=&  \oint dt  \:  j_{ \sigma } (n).   \label{eq: pumpedcharge } 
\eea	

The pumped charge associated with spin $\sigma$ (\ref{eq: pumpedcharge }) 
can be represented as a surface integral rather than a line integral \cite{Brouwer}  
{\small
\bea
 q_{ \sigma }  &=&    \frac{J^{2}}{\pi}  \iint_{ S }  dX_{ -l \sigma } dX_{ l \sigma } \sum_{m = \pm  l} sign(m) \times \notag  \\  
               & &     \partial_{ X_{ m \sigma } }
	              \Im \bigl[ G ( n \sigma  , - m \sigma ;E ) G^{\ast} ( n +1 \sigma  , - m \sigma ;E ) \bigr] |_{E=E_{F}} \notag  \\  
	       & &      \label{eq: surfaceform} 
 \eea}
where $S$ indicates the area which is enclosed by the pumping cycle in parameter space. 
There are two features of the pumped entity (charge or spin),
evident in the last two equations, that we utilize to control the
flow of spin and charge. First, in the line integral form in
Eq.~(\ref{eq: pumpedcharge }) it is clear that reversing the
direction of the time cycle changes the sign of the integral,
indicating a flow in the opposite direction over a pump cycle.
Secondly, the surface integral form, Eq.~(\ref{eq: surfaceform}),
shows that the magnitude of the pumped quantity in a full cycle
depends entirely on the enclosed surface $S$ in parameter space.
Therefore, surfaces that enclose areas with identical functional
form and values of the integrand will yield identical magnitudes
of the pumped charge or spin, while the direction of traversal of
the bounding curve will determine the direction of flow.

\section{ Spin current without charge current }

\begin{figure}
\epsfig{figure=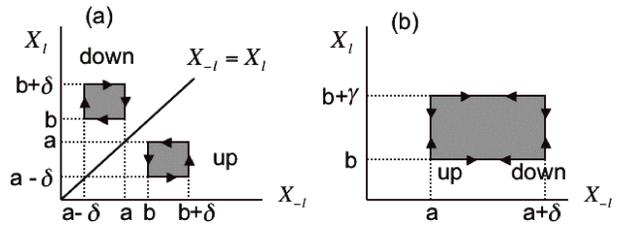,width= 3.2 in}
\caption{(a) Two identical square boxes are located symmetrically
about the $X_{-l}=X_{l}$ line.  For spin up, the parameters $ \{ X_{-l \uparrow}, X_{l \uparrow} \}$
begin at
$(b,a-\delta)$ and go counterclockwise.  For spin down, the $ \{ X_{-l \downarrow}, X_{l \downarrow} \}$
begin at $(a,b)$ and go clockwise.  (b) The spin up
and spin down cycles are located on the same rectangular box.  For spin up,
the cycle begins at $(a,b)$ and goes counterclockwise.  For spin down, the
cycle begins at $(a+\delta,b)$ and goes clockwise.
\label{Fig:cycles}}
\end{figure}

In the previous section we established how we can control the
direction and magnitude of the pumped charge associated with each spin
state, and found expressions to determine them. In particular, varying
the magnetic fields in the generalized parameters allows differential
manipulations of up and down spin states, because the path in
parameter space of the spin-up parameters $\{ X_{-l \uparrow}, X_{l
\uparrow} \}$ becomes distinct from the path in parameter space of the
spin-down parameters $\{ X_{-l \downarrow}, X_{l \downarrow} \}$.  We
will now use these considerations to present two distinct types of
pumping cycles which generate only a pure spin current, with zero
transported charge after each cycle of pumping.  The first cycle
relies on the fact that the integrand of (\ref{eq: surfaceform}) is
symmetric under exchange of $X_{-l \sigma}$ and $X_{l \sigma}$, as one
expects given the form of the Hamiltonian (\ref{eq: finalH}). Consider
two square cycles (we will use the term ``boxes'') of side length
$\delta$ which are located symmetrically in parameter space about the
line $X_{-l} = X_{l}$ (see Fig.\ref{Fig:cycles}a). For the cycle taken
by spin-down parameters $ \{ X_{-l \downarrow}, X_{l \downarrow} \}$,
we pick an arbitrary point $( a, b)_{\downarrow}$ as the initial
choice of parameters. For the cycle taken by spin-up parameters $ \{
X_{-l \uparrow}, X_{l \uparrow } \}$, the initial point $( b,a
-\delta)_{\uparrow}$ is chosen. For those initial points, we find
$u_{-l}/J = (a+b)/2$, $u_{l}/J = (a+b -\delta)/2$ and $E^{Z}_{-l} /J =
(a-b)/2$, $E^{Z}_{l}/J = - (a-b-\delta)/2$. We fix $u_{-l}$,
$E^{Z}_{l}$ throughout the pumping cycle. First, the Zeeman energy at
site $-l$ divided by $J$, $E^{Z}_{-l}/J$, is decreased by $\delta$,
from $(a-b)/2$ to $(a-b -2 \delta)/2$. The resulting motion in
parameter space is parallel to the $X_{-l}$ axis, with the spin-up
parameters moving in the positive direction and the spin-down
parameters moving in the negative direction. Next, the potential
barrier at site $l$ divided by $J$, $u_{l}/J$, is increased by the
amount of $\delta$, from $(a+b- \delta )/2$ to $ (a +b + \delta)
/2$. The spin-down and spin-up parameters both shift upward parallel
to the $X_l$ axis. Next, the Zeeman energy $E^{Z}_{-l}/J$ is increased
by $\delta$, from $(a-b -2 \delta)/2$ back to $(a-b)/2$. Finally, the
potential barrier $u_{l}/J$ is decreased by $\delta$, from $(a+ b +
\delta)/2$ back to $(a +b -\delta)/2$, to complete the cycle. The form
of the definition (\ref{eq: X}) ensures that spin-up and spin-down
parameters shift in opposite directions when the Zeeman energy is
varied and shift in the same direction when the potential barrier,
which results from an electrical potential, is varied. The combination
of these two effects moves the spin-up parameters in a
counterclockwise cycle, and the spin-down parameters in a clockwise
cycle. In addition, our chosen steps generate two square cycles
located symmetrically about the line $X_{-l} = X_{l}$ in parameter
space. Because of the symmetry in (\ref{eq: surfaceform}), these
cycles lead to zero total pumped charge $q_{c} = q_{ \uparrow} +
q_{\downarrow} = 0$. On the other hand, the pumped spin is $q_{s} =
q_{ \uparrow} - q_{\downarrow} \ne 0$ as long as $q_{ \uparrow} ( =
q_{ \downarrow} ) \ne 0$. The result is a pure spin current.


A second type of cycle generates a pure spin current without
relying on the symmetry between $X_l$ and $X_{-l}$. Consider a
rectangular box in parameter space (see Fig.\ref{Fig:cycles}b). By
choosing two initial points appropriately, we can make the
parameters execute cycles on the same rectangular box of width
$\delta$ and height $\gamma$ but in opposite directions. For the
cycle of spin-up parameters, we pick an arbitrary point in
parameter space $(a,b)_{\uparrow}$. For the cycle of spin-down
parameters, the initial point $(a + \delta, b)_{\downarrow}$ is
chosen. These choices correspond to $u_{-l}/J = a + \delta /2$,
$u_{l}/J = b$, $E^{Z}_{-l}/J = \delta /2$, and $E^{Z}_{l}/J = 0$.
We vary $u_{l}$ and $E^{Z}_{-l}$ while fixing $u_{-l}$ and
$E^{Z}_{l}$. First the Zeeman energy $E^{Z}_{-l}/J$ is decreased
by the amount $\delta$, from $\delta /2$ to $  - \delta/2$.
Second, the potential barrier $u_{l}/J$ is increased by $\gamma$,
from $b$ to $b+ \gamma$. Third, the Zeeman energy $E^{Z}_{-l}/J$
is increased by $\delta$, from $- \delta /2$ back to $\delta /2$.
Finally, the potential barrier $u_{l}/J$ is decreased by $\gamma$,
from $b + \gamma$ back to $b$. As a result of these variations,
the spin-up parameters traverse the rectangular box in the
counterclockwise direction while the spin-down parameters traverse
the rectangular box in the clockwise direction. Since the cycles
enclose the same region, but move in opposite directions, a pure
spin current arises.

In this second type of cycle, note that $E^{Z}_{l}$ is fixed at
zero throughout the pumping. This suggests a means of realizing
the cycle experimentally. Rather than trying to produce a
localized magnetic field $B_{-l}$, one could apply a global magnetic field. If all
sites except for the site $-l$ have a negligible g-factor, the
desired Hamiltonian (\ref{eq: finalH}) will arise \cite{Loss}.

We conclude based on the above analysis that, for any given pair of
identical symmetrically located square boxes or for any given single
rectangular box in parameter space, there always exists a pumping
cycle which generates a pure spin current.  This finding implies great
flexibility in the control of pumped pure spin after one cycle. Since
the quantity of pumped spin depends on the shape of the enclosed area
and its location in parameter space, one can tune the quantity of
pumped spin by changing these characteristics of the pumping cycle.

\begin{figure}[t]
\epsfig{figure=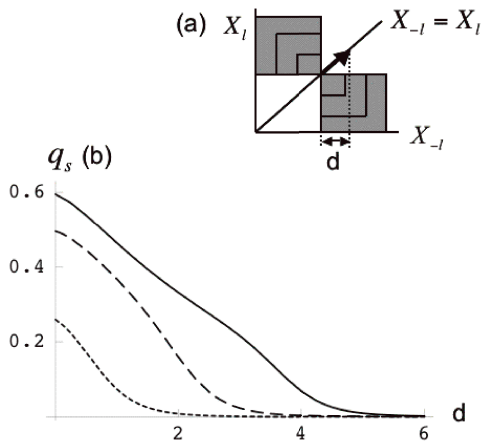, width= 2.5 in}
\caption{ (a) Three pairs of square boxes are located symmetrically
about the median $X_{-l}=X_{l}$ and share the same meeting point.
We move those pairs along the $X_{-l}=X_{l}$ line.
For the three pairs, the box side lengths are 2,4, and 6
respectively. The pairs shift a distance $\sqrt{2}d$ along the $X_{-l}=X_{l}$ line.
(b) Pumped spin $q_{s}$ vs $d$ at fixed $k_{F}=1.4$ and for $l=1$: The solid line is for
the pair of large boxes of side length 6, the long-dashed line is
for the pair of medium boxes of side length 4, and the
short-dashed line is for the pair of small boxes of side length 2.
The result shows monotonic increase of pumped spin with box size
and monotonic decrease of pumped spin with projected distance $d$.
\label{Fig:plotone}}
\end{figure}

\begin{figure}[t]
\epsfig{figure=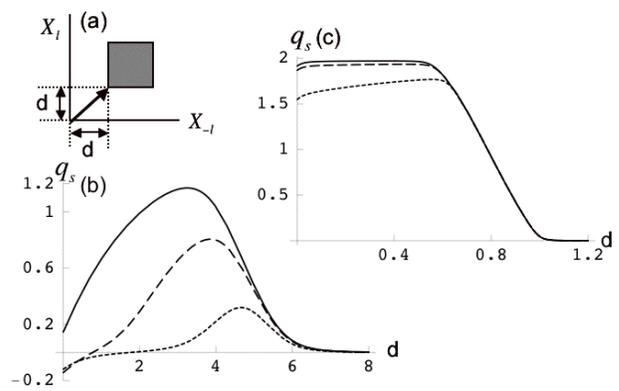,width= 3.2 in}
\caption{Plots of $q_{s}$ vs $d$. 
In each plot, the solid line is for a large box of side length 6, the
long-dashed line is for a medium box of side length 4, and the
short-dashed line is for a small box of side length 2. 
(a) We move each square box along the line $X_{-l}=X_{l}$, where the position of the lower left corner is $(d,d)$.
(b) Plot at $k_{F} = 1.4$ and for $l=1$. For small $d$, negative pumped spin is found for the
smaller two boxes (i.e. spin is pumped in the opposite direction).
(c) Plot at $k_{F}=3.1$ and for $l=1$. The maximum value 2 of pumped spin per
cycle occurs on the solid line. The pumped spin $q_{s}$ is
independent of box size for sufficiently large $d$.
\label{Fig:plottwo}}
\end{figure}

The following plots made using the expressions derived in Sec. II
demonstrate flexibility in controlling a pure spin
current. Fig.~\ref{Fig:plotone} and Fig.~\ref{Fig:plottwo} portray how
the total pumped spin depends on the size of the box in parameter
space and on its location. For the first type of cycle involving pairs
of symmetrically positioned square boxes in parameter space, three
different sizes for the box pairs are considered in
Fig.~\ref{Fig:plotone}a. All meet at the same point on the line
$X_{-l}=X_{l}$. The dependence of the pumped spin on the location of the
cycle in parameter space is studied by moving that common meeting
point a distance $\sqrt{2} d$ along the line $X_{-l}=X_{l}$. Each
curve in Fig.  \ref{Fig:plotone}b shows the variation of the
pumped spin as a function of $d$ for a specific box size, a fixed Fermi
wave vector $k_F = 1.4$ and impurities at $\pm l=\pm 1$.  The
different curves correspond to the three different box sizes. The
plots show a monotonic increase of pumped spin in a cycle as the box
size increases (except where the pumped spin vanishes for all three
box sizes). This is physically reasonable since the box size
determines the difference between the minimum and maximum value of the
potential barriers.  For large boxes, the potential barriers change a
lot during the cycle, resulting in more pumped current, and the
opposite is true for small boxes.  The pumped spin decreases as the
distance $d$ increases. As $d$ increases, the minimum height attained
by the potential barriers gets larger.  As a result, the current must
traverse an increased potential barrier, so that the transmission is
decreased.


For the second type of cycle discussed above for generating a pure
spin current, we consider three square boxes that each straddle the
median line $X_{-l}=X_{l}$ as shown in Fig.~\ref{Fig:plottwo}a. The
pumped spin in a cycle is plotted as a function of $d$ where the lower
left corner of the box is at the point $(d,d)$.  The two plots,
Fig.~\ref{Fig:plottwo}b and Fig.~\ref{Fig:plottwo}c, correspond to
different choices of Fermi wave vector, $k_{F} = 1.4$ and $k_{F} =
3.1$ respectively; the clear differences between the two sets of
curves indicate that the Fermi wave vector is another essential factor
in controlling the spin current flow. For both plots the impurity
locations are $\pm l = \pm 1$.  In Fig.~\ref{Fig:plottwo}b, the pumped
spin increases monotonically with box size at most locations $d$.  For
large $d$, the behavior of the curves in Fig.~\ref{Fig:plottwo}b shows
monotonic decrease with $d$ like the curves in
Fig.~\ref{Fig:plotone}b. However, for small $d$ the pumped spin
increases with $d$, so that each curve in Fig.~\ref{Fig:plottwo}b has
maximum pumped spin between $d=3$ and $d=5$. This kind of maximum can be explained
in terms of resonant transmission\cite{Levinson}. When the pumping
cycle includes locations in parameter space for which the Fermi energy
satisfies a resonance condition, an enhanced transmission coefficient
leads to a large pumped current.  For the parameters chosen in the
figure, a line of resonant points in parameter space runs near
$(X_l,X_{-l}) = (6,6)$.  As each box shifts with increasing $d$ to
enclose these resonant points, the pumped current grows 
even though the minimum heights of the potential barriers get larger.  
For $d$ near zero in the case of the two smaller
boxes in Fig.~\ref{Fig:plottwo}b, there is actually a region of negative values for the pumped
spin, indicating a reversal of direction of the spin flow.  The
direction of pumped spin changes as the box moves or its size
increases. This shows we can control the direction of the spin current
without reversing the whole cycle, by simply adjusting the box size or
location.

For Fig.~\ref{Fig:plottwo}c, the maximum pumped
spin for the largest box is $2$, so that during each cycle
one spin-up carrier goes to the right and one spin-down carrier goes
to the left.  This large maximum value is due to both the low minimum barrier heights 
and the resonance transmission. For $k_{F}=3.1$, a resonant line runs near
$(X_l,X_{-l}) = (1,1)$ and boxes enclosing the resonant line also have the low minimum
barrier heights, by adding up two effects, the pumped spin has the large maximum value such as $2$. 
With increasing $d$ from the origin, the pumped spin
decreases to assume the same finite value for the three different box
sizes, indicating that there is little variation in the integrand in
Eq.~(\ref{eq: surfaceform}) in the surface integral far from the
origin whereas there are stronger variations close to the origin.


\begin{figure}
\epsfig{figure=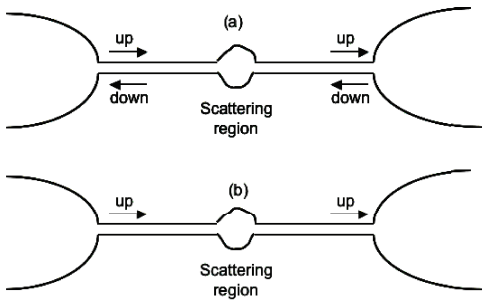,width= 2.5 in}
\caption{ (a) An example of pure spin current. There is no net flow of charge.
(b) An example of selective spin current. Only one kind of spin contributes to current, 
there is no cancellation of charge. \label{compare}}
\end{figure}


\section{ Selective spin pumping }

\begin{figure}[t]
\epsfig{figure=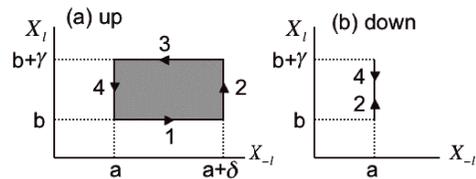,width= 2.5 in}
\caption{ (a) Pumping cycle for spin-up parameters, beginning at
the point (a,b) in the generalized parameter space and proceeding
counterclockwise. It encloses a non-zero area, and generates a
non-zero spin-up current. (b) Pumping cycle for spin-down
parameters, also beginning at (a,b). This cycle encloses no
area, so there is no spin-down current generated.
\label{Fig:selectivepumping}}
\end{figure}

We consider another type of cycle which has a spin-filtering effect. 
This cycle selectively pumps one kind of spin, 
so that an equal spin and charge current flow (See Fig. \ref{compare}). 
Consider a rectangular box in parameter space. We choose the same initial
point $(a,b)_{\uparrow \downarrow}$ for both spin-up and spin-down
parameters. This choice of initial point implies that $u_{-l}/J
=a, u_{l}/J =b$ and $E^{Z}_{-l} = E^{Z}_{l}=0$.  We fix
$E^{Z}_{l}$ at zero. First, we increase $u_{-l}$ and decrease
$E^{Z}_{-l}$ simultaneously in such a fashion that
$(u_{-l}+E^{Z}_{-l})/J$ remains at the initial value $a$ while
$(u_{-l}-E^{Z}_{-l})/J$ is increased by the amount $\delta$, from
$a$ to $a+\delta$. At the end of this process, $u_{-l}/J$ is
$a+\delta/2$ and $E^{Z}_{-l}/J$ is $- \delta/2$. The spin-up
parameters shift parallel to $X_{-l}$, but the spin-down
parameters remain unchanged. Second, $u_{l}/J$ is increased by
$\gamma$, from $b$ to $b+\gamma$. The spin-down parameters and the
spin-up parameters both shift parallel to the $X_{l}$ axis. Next,
we decrease $u_{-l}$ and increase $E^{Z}_{-l}$ simultaneously,
keeping $u_{-l}+E^{Z}_{-l}$ fixed while $(u_{l}-E^{Z}_{l})/J$
decreases by $\delta$, from $a+\delta$ back to $a$. This produces
a path parallel to $X_{-l}$ for spin-up parameters, but does not
shift the spin-down parameters at all. Finally, $u_{l}/J$ is
decreased by $\gamma$, from $b+\gamma$ back to $b$, shifting both
spin-down parameters and spin-up parameters parallel to $X_{l}$.
For spin-up parameters, this cycle makes rectangular path
enclosing a non-zero area (see Fig. \ref{Fig:selectivepumping}a)
that produces a current of spin-up carriers.  On the other hand,
the cycle for spin-down shifts along a straight line in parameter
space that encloses no area (see Fig. \ref{Fig:selectivepumping}b)
and pumps no current. The result is perfect selective spin pumping of
spin-up carriers.  Naturally, a selective current of spin-down
carriers can be generated with trivial modifications to this
protocol.


More generally, we can transfer charge and spin to achieve any
rational value of $q_{s}/q_{c}$ by combining and repeating
spin selective cycles. Suppose that the value
$q_{s}/q_{c} = M/N$ is desired, where $M$ and $N$ are integers. It
is always possible to find two integers $n$ and $m$ satisfying
$n/m = (N-M)/(N+M)$. By performing $|m|$ selective spin-up cycles
and $|n|$ selective spin-down cycles, we can generate an arbitrary
rational value for $q_{s}/q_{c}$. (If $m$ is positive, the spin-up
cycles should be traversed in a counterclockwise direction, while
for negative $m$ they should be traversed in a clockwise
direction.  The same is true for $n$ and the spin-down cycles.)

\section{ Arbitrary combinations of spin current and charge current }

In earlier sections, we gave definite cycles that could be used to
pump spin with no charge, to selectively pump carriers of a given
spin orientation, or to pump a rational ratio of spin to charge.
Here, we argue that other cycles can produce arbitrary ratios of
spin current to charge current, requiring only one cycle of
pumping with no repetition.  (However, we do not give a recipe for
identifying the cycle; trial and error tuning may be needed.)


\begin{figure}[t]
\epsfig{figure=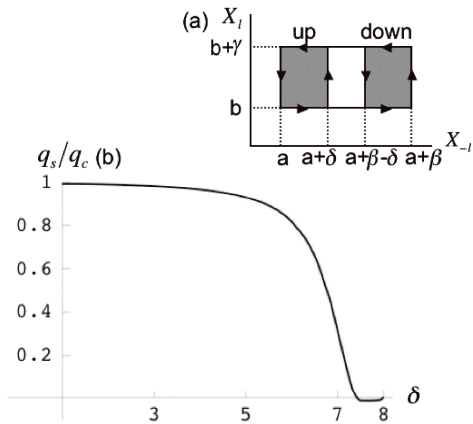,width= 2.5 in}
\caption{ (a) We consider two congruent rectangular boxes
symmetrically located within one large rectangular box. By varying
$\delta$, we shift the interior edges of the rectangular boxes.
The left rectangular box shows the trajectory of spin-up pumping
parameters, while the right rectangular box is for spin-down
pumping parameters. For the first type of cycle (shown here), both the
small rectangles are traversed in a counterclockwise direction.
For the second type of cycle (not shown here), spin-up parameters
traverse in a counterclockwise direction while spin-down
parameters traverse in a clockwise direction. (b) Plot of
$q_{s}/q_{c}$ vs $\delta$ for first type of cycle at $k_{F}=3.1$ and for $l=1$. 
We set $(a,b) = (0,0)$ and $\beta = \gamma = 8$.
$\delta$ is varied from $1$ to $8$.
Any $q_{s}/q_{c}$ in the range $0$ to $1$ can be found. 
Note that the origin of plot is $(1,0)$, not $(0,0)$.
\label{Fig:ratioplot}}
\end{figure}


Consider a single rectangular box and put two congruent rectangular
cycles symmetrically at its ends as shown in
Fig. \ref{Fig:ratioplot}a.  The left cycle is traversed by the spin-up
parameters, and the right cycle by the spin-down parameters.  We first
describe a protocol in which the spin-up and spin-down parameter
cycles are both traversed in a counterclockwise direction.  Choose
initial points at the lower left corner of each cycle.  For
spin-up parameters, the point is $(a,b)_{\uparrow}$.  For spin-down
parameters, the initial point is $(a+\beta-\delta,b)_{\downarrow}$.
These choices imply that the physical parameters are $u_{-l}/J =a+ (
\beta-\delta)/2$, $u_{l}/J=b$, and $E^{Z}_{-l}/J = (\beta-\delta)/ 2$,
and $E^{Z}_{l} =0$.  Suppose that the Zeeman energies $E^{Z}_{-l}$ and
$E^{Z}_{l}$ are fixed.  As a first step, we increase $u_{-l}/J $ by
$\delta$, from $a+ ( \beta-\delta)/2$ to $a+ ( \beta+\delta)/2$.
Next, we increase $u_{l}/J$ by $\gamma$, from $b$ to $b+ \gamma$.
Then, we decrease $u_{-l}/J$ by $\delta$, from $a+ ( \beta+\delta)/2$
back to $a+ ( \beta-\delta)/2$.  Finally, we decrease $u_{l}/J$ by
$\gamma$, from $b+ \gamma$ back to $b$.  This is the complete
protocol.  For counterclockwise traversal, it is typically the case
that the charge $q_{\uparrow}$ produced by the spin-up cycle will be
positive and so will the charge $q_{\downarrow}$ produced by the
spin-down cycle.  As a result, the ratio $q_{s}/q_{c} = (
q_{\uparrow}- q_{\downarrow} ) /( q_{\uparrow} + q_{\downarrow} )$
will typically satisfy $ |q_{s}/q_{c}| < 1 $.  Tuning the parameters
will typically permit any desired value of the ratio, as shown in
Fig.\ \ref{Fig:ratioplot}b.  In order to attain a ratio $ |q_{s}/q_{c}|
> 1 $, one can use the same two rectangular cycles, traversed in, say,
a counterclockwise direction for spin-up and in a clockwise direction for
spin-down.  The lower left point in the spin-up rectangle and the
lower right point in the spin-down rectangle serve as initial
points. It follows that the physical parameters take the initial
values $u_{-l}/J= a+\beta/2$, $u_{l}/J= b$, $E^{Z}_{-l}/J=
\beta/2$, and $E^{Z}_{l}= 0$.  We fix $u_{-l}$, $E^{Z}_{l}$ during
the cycle. First, $E^{Z}_{-l}/J$ is decreased by $\delta$, from
$\beta/2$ to $\beta/2-\delta$.  Next, $u_{l}/J$ is increased by
$\gamma$, from $b$ to $b + \gamma$.  Third, $E^{Z}_{-l}/J$ is
increased by $\delta$, from $\beta/2-\delta$ back to $\beta/2$.
Finally, $u_{l}/J$ is decreased by $\gamma$, from $b + \gamma$
back to $b$.  Since we are traversing the same two rectangles as
the previous protocol, we see that the same value of
$q_{\uparrow}$ will be produced, but the spin-down charge will now
be $-q_{\downarrow}$, where $q_{\downarrow}$ is defined as the
spin-down charge produce by a counterclockwise traversal.  The
result is that $q_{s}/q_{c} = ( q_{\uparrow} + q_{\downarrow} ) /(
q_{\uparrow}- q_{\downarrow} )$, which is simply the inverse of
the value of the ratio obtained in the first protocol, so that now $
|q_{s}/q_{c}| > 1 $ typically.  Given these two protocols, we
should be able to achieve arbitrary combinations of spin current
and charge current over the whole range $0 \le |q_{s}/q_{c}| \le
+\infty$ by varying $\delta$ (see Fig.\ref{Fig:ratioplot}b) 
which determines each pumping cycle within the rectangular box.
Although tuning is required to achieve a given arbitrary ratio,
the exact inverse ratio for that combination can be attained
predictably by reversing one of cycles. The extreme cases
$|q_{s}/q_{c}|= + \infty$ and $|q_{s}/q_{c}|=1$ correspond
respectively to the cases of pure spin pumping and to spin
selective pumping described above.  If $q_{s}/q_{c}$ is positive,
a corresponding negative ratio, which has the same absolute value,
can be obtained by exchanging cycles so that the left box is for
the spin down parameters and the right box is for the spin up
parameters. Thus, all possible ratios $q_s/q_c$ are attainable.

\section{Conclusion}

In conclusion, we have presented a deterministic way to produce a
pure spin current, a spin selective current, or a rational ratio
of spin to charge current. Our proposal relies on generalized
pumping parameters, each of which depends on more than one
physical parameter.  In our calculations, the maximum value 2
for the pumped spin is observed for some pumping cycles, and we
find that the direction of the spin current can be manipulated via
the size or location of the pumping cycle in parameter space.   We
also presented an argument to show that it is typically possible
in a single cycle to pump an arbitrary ratio of spin current to
charge current (although some trial and error may be needed to
find the right cycle).  These results suggest that adiabatic
quantum pumping could be a versatile tool for generating a desired
current in a spintronics device.

\acknowledgments

The authors gratefully acknowledge the support of the Packard Foundation and of NSF NIRT program grant DMR-0103068.



\begin{thebibliography}{}
\bibitem{Samarth}
  {\em Semiconductor Spintronics and Quantum Computation}, edited by D.D. Awschalom, D. Loss, and N. Samarth (Springer, 2002).
\bibitem{Brouwer}
  P.W. Brouwer, Phys. Rev. B {\bf 58}, R10 135 (1998).
\bibitem{Avron}
  J.E. Avron, A. Elgart, G.M. Graf, and L. Sadun, Phys. Rev. B {\bf 62}, R10 618 (2000).
\bibitem{Entin}
  O. Entin-Wohlman, A. Aharony, and Y. Levinson, Phys. Rev. B {\bf 65}, 195411 (2002).
\bibitem{Buttiker}
  M. Moskalets and M. B\"uttiker, Phys. Rev. B {\bf 66}, 205320 (2002).
\bibitem{Sharma}
  P. Sharma and C. Chamon, Phys. Rev. Lett. {\bf 87}, 096401 (2001).
\bibitem{Mucciolo}
  E.R. Mucciolo, C. Chamon, and C.M. Marcus, Phys. Rev. Lett. {\bf 89}, 146802 (2002).
\bibitem{Watson}
  S.K. Watson, R.M. Potok, C.M. Marcus, and V. Umansky, Phys. Rev. Lett. {\bf 91}, 258301 (2003).
\bibitem{Aono}
  T. Aono, Phys. Rev. B {\bf 67}, 155303 (2003).
\bibitem{Wei}
  Y. Wei,  L. Wan, B. Wang, and J. Wang, Phys. Rev. B {\bf 70}, 045418 (2004).
\bibitem{Economou}
  E.N. Economou,  {\em Green's Functions in Quantum Physics}, (Springer-Verlag, 1979).
\bibitem{Buttiker-Landauer}  
  M. B\"{u}ttiker and R. Landauer, Phys. Rev. Lett. {\bf 49}, 1739 (1982).
\bibitem{Liliana}
  L. Arrachea, Phys. Rev. B {\bf 72}, 125349 (2005). 
\bibitem{Loss}
  P. Recher, E.V. Sukhorukov, and D. Loss, Phys. Rev. Lett. {\bf 85}, 1962 (2000).
\bibitem{Levinson}
  Y. Levinson, O. Entin-Wohlman, and P. W\"olfle, Physica A {\bf 302}, 335 (2001).
\end{thebibliography}
\end{document}